\documentclass[aps,prd,twocolumn,superscriptaddress,nofootinbib]{revtex4-1}

\usepackage{amsmath,amssymb}

\usepackage{graphicx}
\usepackage{dcolumn}

\begin{document}

\title{Constraining interacting dark energy with CMB and BAO future surveys}

\author{Larissa Santos}
\email{larissa@ustc.edu.cn}
\affiliation{CAS Key Laboratory for Researches in Galaxies and Cosmology, Department of Astronomy, University of Science and Technology of China, Chinese Academy of Sciences, Hefei, Anhui 230026, China, and\\
School of Astronomy and Space Science, University of Science and
Technology of China, Hefei 230026, China}

\author{Wen Zhao}
\affiliation{CAS Key Laboratory for Researches in Galaxies and Cosmology, Department of Astronomy, University of Science and Technology of China, Chinese Academy of Sciences, Hefei, Anhui 230026, China, and\\
School of Astronomy and Space Science, University of Science and
Technology of China, Hefei 230026, China}

\author{Elisa G.~M.~Ferreira}
\affiliation{Department of Physics, McGill University, Montr\'{e}al, QC, H3A 2T8, Canada}

\author{Jerome Quintin}
\thanks{Vanier Canada Graduate Scholar.}
\affiliation{Department of Physics, McGill University, Montr\'{e}al, QC, H3A 2T8, Canada}

\date{\today}

\begin{abstract}
In this paper, we perform a forecast analysis to test the capacity of future baryon acoustic oscillation (BAO) and cosmic microwave background (CMB) experiments to constrain phenomenological interacting dark energy models using the Fisher matrix formalism. We consider a Euclid-like experiment, in which BAO measurements is one of the main goals, to constrain the cosmological parameters of alternative cosmological models. Moreover, additional experimental probes can more efficiently provide information on the parameters forecast, justifying also the inclusion in the analysis of a future ground-based CMB experiment mainly designed to measure the polarization signal with high precision. In the interacting dark energy scenario, a coupling between dark matter and dark energy modifies the conservation equations such that the fluid equations for both constituents are conserved as the total energy density of the dark sector. In this context, we consider three phenomenological models which have been deeply investigated in literature over the past years. We find that the combination of both CMB and BAO information can break degeneracies among the dark sector parameters for all three models, although to different extents. We found powerful constraints on, for example, the coupling constant when comparing it with present limits for two of the models, and their future statistical 3-$\sigma$ bounds could potentially exclude the null interaction for the combination of probes that is considered. However, for one of the models, the constraint on the coupling parameter does not improve the present result (achieved using a large combination of surveys), and a larger combination of probes appears to be necessary to eventually claim whether or not interaction is favored in that context.
\end{abstract}


\maketitle

\section{Introduction\label{Introduction}}

Cosmic microwave background (CMB) measurements have contributed to an unprecedented understanding of the universe at its early stages,
leading us to a well-established and consistent picture of how the universe is today. Recent results from the Planck satellite collaboration
showed that baryonic matter constitutes about $5 \%$ of all that is known, leaving about $95 \%$ of `dark' components
\cite{Ade:2013zuv,Ade:2015xua}. The cold dark matter (CDM) accounts for roughly $26\%$ of the universe, and the remaining
$69\%$ is in the form of dark energy (DE).
Considering the standard cosmological model, the DE assumes its simplest form as a cosmological constant $\Lambda$, leading
to the so-called $\Lambda$CDM model. Based on general relativity, the cosmological constant viewed as a DE fluid with equation of
state (EoS) $w_\Lambda=p_\Lambda/\rho_\Lambda=-1$, where $p_\Lambda$ is the pressure and $\rho_\Lambda$
the energy density, can explain well the current accelerated expansion of the universe \cite{Riessetal,Perlmutter:1998np}.

Despite of successfully explaining the observations, the
$\Lambda$CDM model faces some difficulties \cite{Bull:2015stt}, especially in the dark
sector. Dark matter (DM) particles have not been detected yet, and
their origin, presumably beyond the standard model of particles
physics, is still unknown (see, e.g., Ref.~\cite{DMref} and
references therein). Theoretical calculations of the vacuum energy
density estimate the value of the cosmological constant to be
orders of magnitude larger than its actual observed value (see,
e.g., Ref.~\cite{DEref}). In addition, the present values of the
DM and DE densities are of the same order of magnitude even though
they do not share the same cosmological evolutionary behaviour.
This cosmic coincidence seems to indicate that we are living in a
special epoch of the cosmic history \cite{Zlatev:1998tr}.
To overcome some of these problems, researchers
started to consider models in which DM and DE interact,
and these became very useful in alleviating this coincidence
problem (see, e.g.,
Refs.~\cite{coincidenceproblemrefothers,coincidenceproblemrefPavon,coincidenceproblemrefdelCampoetal,Wang:2005jx,coincidenceproblemrefZhao,He:2010im,Costa:2013sva,Wang:2016lxa}).
An interacting DM and DE scenario would affect the overall
evolution of the universe and its expansion history, thus it is
observationally distinguishable from the $\Lambda$CDM model. The
interaction can then be constrained by the data, becoming a
testable theory for the universe.

Present observations, however, are not able to confidently distinguish between these alternative interacting DE models
and $\Lambda$CDM. Updated cosmological data have already been confronted with such models (see, for example,
Refs.~\cite{Costa:2013sva,Wang:2016lxa,Murgia:2016ccp,Costa:2016tpb,iDEobsrefPettorino,iDEobsrefSalvatelliMena,iDEobsrefYangXu,iDEobsrefmoreYang,iDEobsrefLi,iDEobsrefNunes,Marcondes:2016reb,An:2017kqu,iDEobsrefothers}),
but often a null interaction cannot be discarded with high confidence.
Nevertheless, Ref.~\cite{Abdalla:2014cla} claims that interacting models can explain the high-redshift observations
of the Baryon Oscillation Spectroscopic Survey (BOSS) \cite{Delubac:2014aqe}, which deviate from $\Lambda$CDM. In this context, future generation of
astronomical \mbox{ground- and} space-based experiments as well as future CMB experiments will be able to precisely perform consistency tests
of the $\Lambda$CDM model and significantly improve constraints on alternative scenarios, including the interacting DE models.
A lot of effort has been made to constrain and forecast parameters in alternative DE scenarios in the past years (see,
for instance, Refs.~\cite{forecastrefother,Zhao:2010sz,Santos:2013gqa,forecastrefMartinelli}).
However, for interacting DE models, only recently,
Ref.~\cite{Caprini:2016qxs} performed a forecast analysis of the capability of eLISA to constraint such models, finding that it can
only be competitive if the onset of the deviation from $\Lambda$CDM of these models occurs relatively late in the evolutionary history of the universe.
Earlier studies are outdated since they have explored the forecast of Planck-like CMB surveys alone on phenomenological
interacting DE models \cite{Martinelli:2010rt} or with earlier configurations of Euclid-like experiments \cite{DeBernardis:2011iw}.
Others have explored the forecast for field theory implementations of coupled DE \cite{Amendola:2011ie}.
In this paper, we consider a combination of future state-of-the-art probes:
the baryon acoustic oscillations (BAO) information that can be obtained from an updated
Euclid-like experiment \cite{Amendola:2012ys} and the primary CMB fluctuations from a possible future experiment like AdvACT \cite{Calabrese:2014gwa}.
The goal is to test their ability to constrain the phenomenological interacting DE models described in this paper
and determine how their combination can help break the degeneracies between the different cosmological parameters.

The paper is organized as follows. In Sec.~\ref{models},
we describe the phenomenological models with which we perform the parameter forecast.
Sec.~\ref{Method} is devoted to the methodology we use to calculate the marginalized errors on the chosen parameters,
followed by the results in Sec.~\ref{Results}. Finally, in Sec.~\ref{conclusions}, we draw our conclusions.

\section{The interacting dark energy models\label{models}}

In the standard cosmological model, the energy momentum tensor for radiation, baryons, cold DM, and DE is conserved separately,
i.e., for each component. Conversely, in an interacting DE model, the fluid equations for the DE and DM are not
conserved individually, but the dark sector as a whole satisfies the usual energy-conservation equation.
In a Friedmann-Robertson-Walker universe, the conservation equations for the fluids that exchange energy are:
\begin{align}
\dot{\rho}_\mathrm{DM} + 3H\rho_\mathrm{DM} &=+Q\,, \nonumber \\
\dot{\rho}_\mathrm{DE} + 3H(1+ w_\mathrm{DE})\rho_\mathrm{DE} &=-Q\,,
\label{de_dm}
\end{align}
where $H$ is the Hubble parameter, ${\rho}_\mathrm{DM}$ and ${\rho}_\mathrm{DE}$ are the energy densities for DM and DE,
respectively, and $w_\mathrm{DE}\equiv p_\mathrm{DE}/\rho_\mathrm{DE}$ is the DE EoS.
It is clear from the DM conservation equation that we assume $p_\mathrm{DM}=0$, i.e., the DM EoS is that of pressureless matter (dust).
Here, $Q$ represents the interaction kernel that can be written
phenomenologically as $Q= 3H(\xi_1\rho_\mathrm{DM} + \xi_2\rho_\mathrm{DE})$, where the coupling coefficients
(the constants $\xi_1$ and $\xi_2$) are to be determined by observations (see, e.g.,
Refs.~\cite{He:2010im,Costa:2013sva,Abdalla:2014cla,Costa:2016tpb,Wang:2016lxa,phenoref,He:2008tn}).
The energy flow from DE to DM is defined by $Q > 0$, and conversely,
$Q < 0$ defines an energy flow from DM to DE. Considering the stability of the cosmological perturbations
when $w_\mathrm{DE}$ is kept constant, two choices can be made \cite{He:2008si}:
first, one can take $\xi_1 = 0$ and $\xi_2 \neq 0$, together with a constant DE EoS within the range $-1 < w_\mathrm{DE} < -1/3$
(dubbed model 1) or $w_\mathrm{DE} < -1$ (model 2); second, one can take $\xi_2 = 0$ and $\xi_1 \neq 0$ with $w_\mathrm{DE} < -1$,
defining our third considered model (for a summary, see Table \ref{ide_m}).
For all three models, the other components follow the standard conservation equations.
For a review of the topic, we refer to Refs.~\cite{Bolotin:2013jpa,Wang:2016lxa}.

\begin{table}[ht]
\centering \caption{Interacting DE models considered in the analysis of this paper.\label{ide_m}}
\begin{ruledtabular}
\begin{tabular}{ccc}
    Model & $Q$ & DE EoS  \\
    \hline
    1 & $3\xi _{2} H\rho_\mathrm{DE}$ & $-1 <  w_\mathrm{DE} < -1/3$  \\
    2 & $3\xi_{2}H\rho_\mathrm{DE}$  & $w_\mathrm{DE} < -1$ \\
    3 & $3\xi_{1}H\rho_\mathrm{DM}$  & $w_\mathrm{DE} < -1$
\end{tabular}
\end{ruledtabular}
\end{table}

When one allows for an energy flow between DE and DM, the energy densities present a different evolution for each model.
The presence of this change in the redshift dependency leads to an effective EoS for DM and for DE,
which depends on the form of the interaction.
For models 1 and 2, the energy densities for DM and DE are given by (see Ref.~\cite{He:2008tn})
\begin{align}
\rho_\mathrm{DE}=&~(1+z)^{3\left(1+w_\mathrm{DE}+\xi_{2}\right)}\rho_\mathrm{DE}^0\,, \nonumber \\
\rho_\mathrm{DM}=&~(1+z)^3 \nonumber \\
  &\times\left\{\frac{\xi_{2}\left[1-(1+z)^{3(\xi_{2}+w_\mathrm{DE})}\right]\rho_\mathrm{DE}^0}{\xi_{2}+w_\mathrm{DE}}+\rho_\mathrm{DM}^0\right\}\,,
\label{eq:rho12}
\end{align}
and the effective equations of state are
\begin{align}
w_\mathrm{DE}^\mathrm{eff} = w_\mathrm{DE} + \xi_2 \,, \qquad w_\mathrm{DM}^\mathrm{eff} = -\xi_2/r\,,
\label{eff_EoS_model12}
\end{align}
with $r\equiv\rho_\mathrm{DM}/\rho_\mathrm{DE}$.
For model 3, the evolution of the energy densities follows
\begin{align}
\rho_\mathrm{DE}=&~(1+z)^{3(1+w_\mathrm{DE})}\left(\rho_\mathrm{DE}^{0}+\frac{\xi_{1}\rho_\mathrm{DM}^{0}}{\xi_{1}+w_\mathrm{DE}}\right) \nonumber \\
&-\frac{\xi_{1}}{\xi_{1}+w_\mathrm{DE}}(1+z)^{3(1-\xi_{1})} \rho_\mathrm{DM}^{0}~, \nonumber \\
\rho_\mathrm{DM}=&~\rho_\mathrm{DM}^0 (1+z)^{3-3 \xi_{1}}\,,
\label{eq:rho3}
\end{align}
and the effective equations of state are
\begin{align}
w_\mathrm{DE}^\mathrm{eff} = w_\mathrm{DE} + \xi_1r \,, \qquad w_\mathrm{DM}^\mathrm{eff} = -\xi_1\,.
\label{eff_EoS_model3}
\end{align}
In both cases, the baryon energy density ($\rho_\mathrm{b}$) is given by the standard expression, i.e., it is proportional to $(1+z)^3$.
Note that the quantities measured today are identified by the superscript $0$.
For example, using the definition of the cold DM density parameter today, $\Omega_\mathrm{c}\equiv\rho_\mathrm{DM}^0/\rho_\mathrm{crit}$,
where $\rho_\mathrm{crit}\equiv 3H_0^2/(8\pi G)$ is the critical density of the universe, one has
\begin{equation}
 \rho_\mathrm{DM}^0=\frac{3\omega_\mathrm{c}}{8\pi G}\times (100~\mathrm{km~s^{-1}~Mpc^{-1}})^2\,,
\end{equation}
where we use the same notation as in Ref.~\cite{Ade:2013zuv} with
$h$ defined such that $H_0=100\,h~\mathrm{km~s^{-1}~Mpc^{-1}}$,
and where $H_0$ and $\omega_{c}\equiv h^2\Omega_\mathrm{c}$ are the Hubble parameter and the physical density of cold DM today.

Since DM and DE are currently only measured gravitationally and since gravity only probes the total energy momentum tensor,
degeneracies in the cosmological parameters are inevitable. As it is already known in the literature
(see, e.g., Refs.~\cite{Kunz:2007rk,He:2010im,Ade:2015rim}), and as we can see
in the expressions for the energy densities of the coupled dark components, there is a degeneracy between $w_\mathrm{DE}$ and $\Omega_\mathrm{c}$.
At the background level, the fact that only the total energy momentum can be measured also leads to a degeneracy between
the coupling constant and $w_\mathrm{DE}$, as we can see in the effective DE EoS for models 1 and 2 [see Eq.~\eqref{eff_EoS_model12}].
For model 3, this degeneracy is no longer present today since $w_\mathrm{DE}^\mathrm{eff}\simeq w_\mathrm{DE}$
for $r\ll 1$ (i.e., when $\rho_\mathrm{DE}\gg\rho_\mathrm{DM}$).
In that case, the DE EoS and the interacting constant can be measured independently using the background evolution \cite{He:2010im}.

\begin{figure}[ht]
{\includegraphics[scale=0.8]{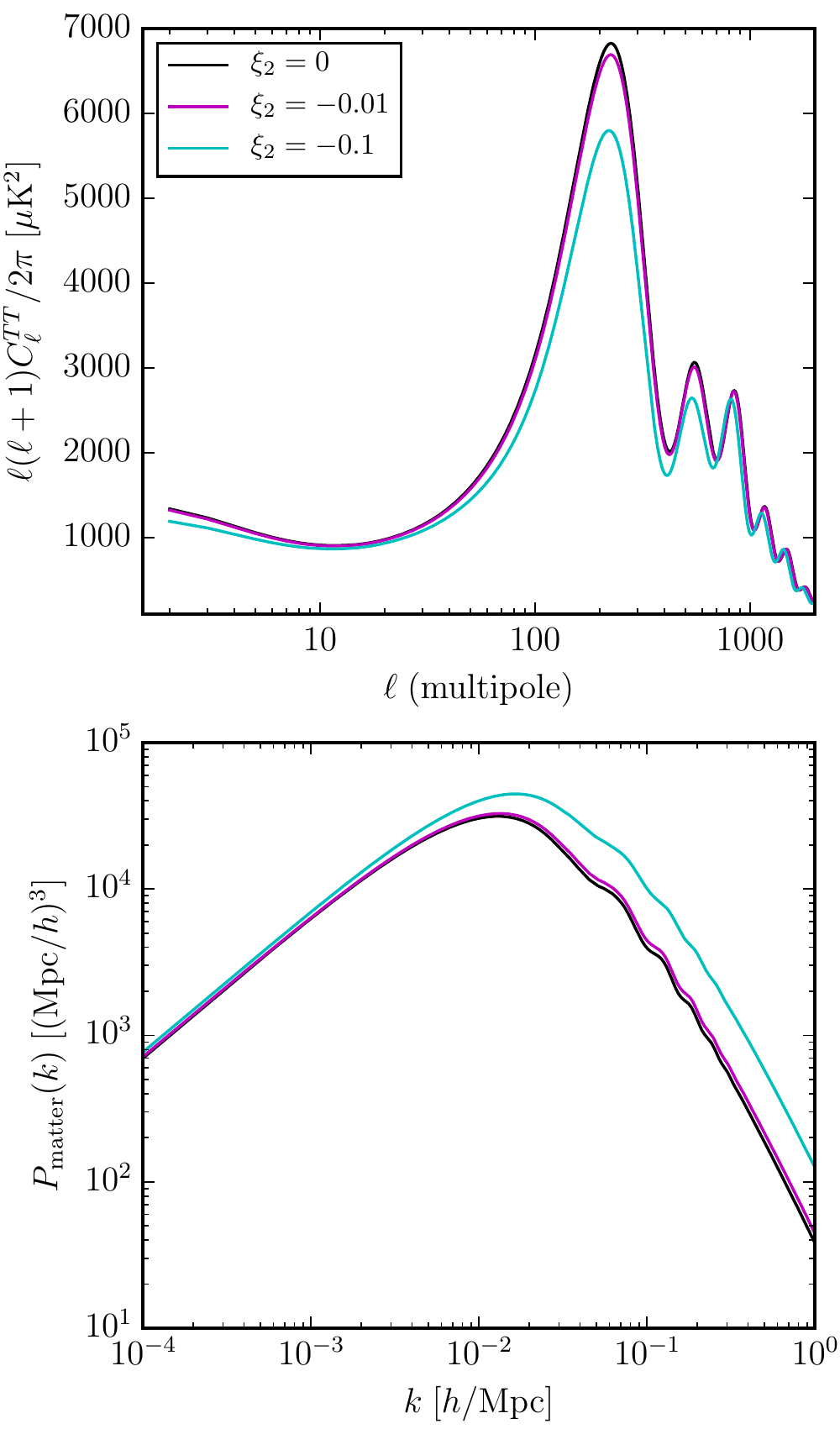}}
 \caption{Plots of the CMB temperature power spectrum (upper panel) and matter power spectrum (lower panel)
 for three different coupling parameters $\xi_2$ for model 1.
 The other parameters assume the fiducial values given in Sec.~\ref{Method}, Table \ref{tbl_m1}.
 In particular, $w_\mathrm{DE}=-0.9434$.}
\label{model1}
\end{figure}

\begin{figure}[ht]
{\includegraphics[scale=0.8]{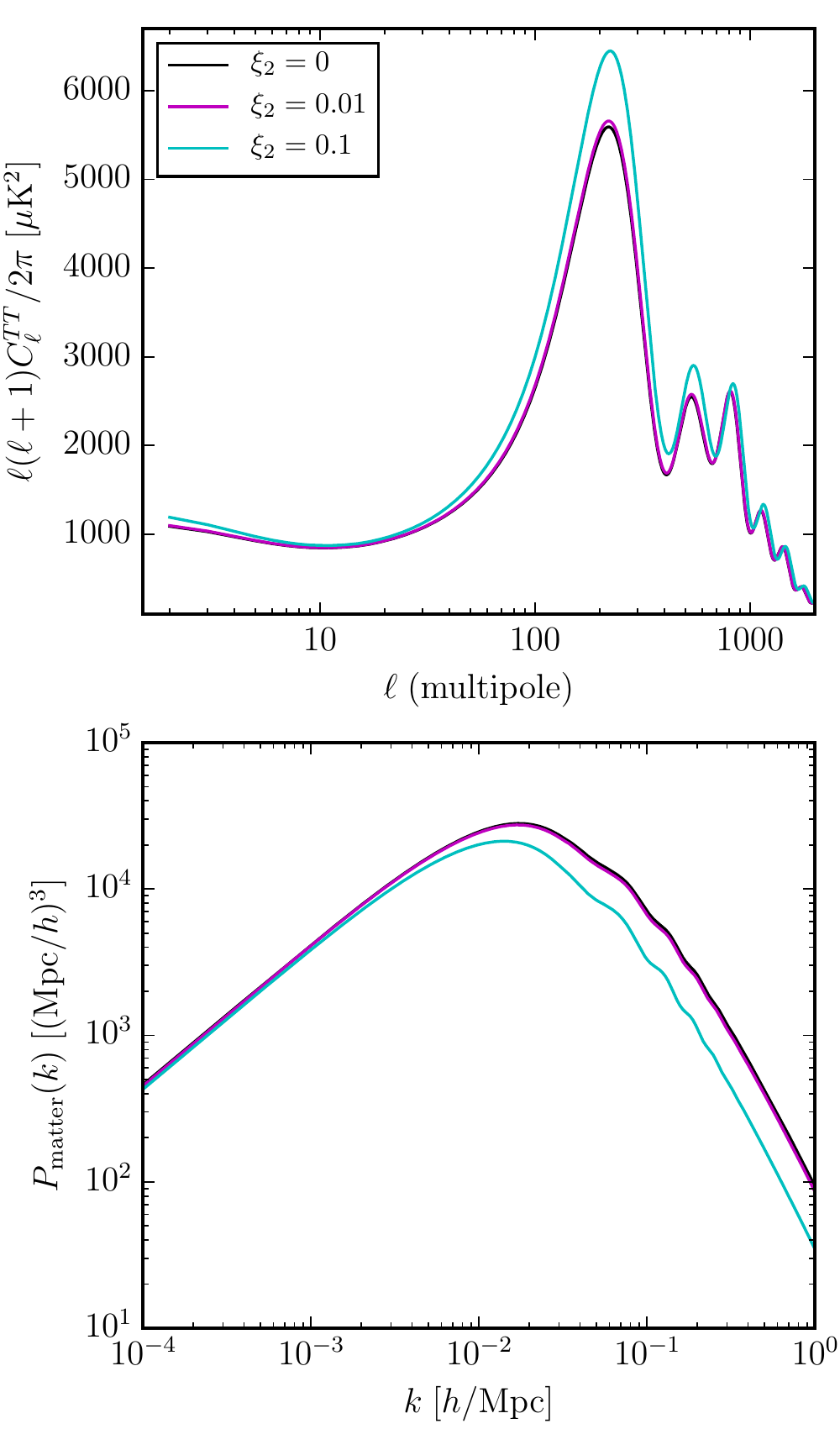}}
 \caption{Plots of the CMB temperature power spectrum (upper panel) and matter power spectrum (lower panel)
 for three different coupling parameters $\xi_2$ for model 2.
 The other parameters assume the fiducial values given in Sec.~\ref{Method}, Table \ref{tbl_m2}.
 In particular, $w_\mathrm{DE}=-1.087$.}
\label{model2}
\end{figure}

\begin{figure}[ht]
{\includegraphics[scale=0.8]{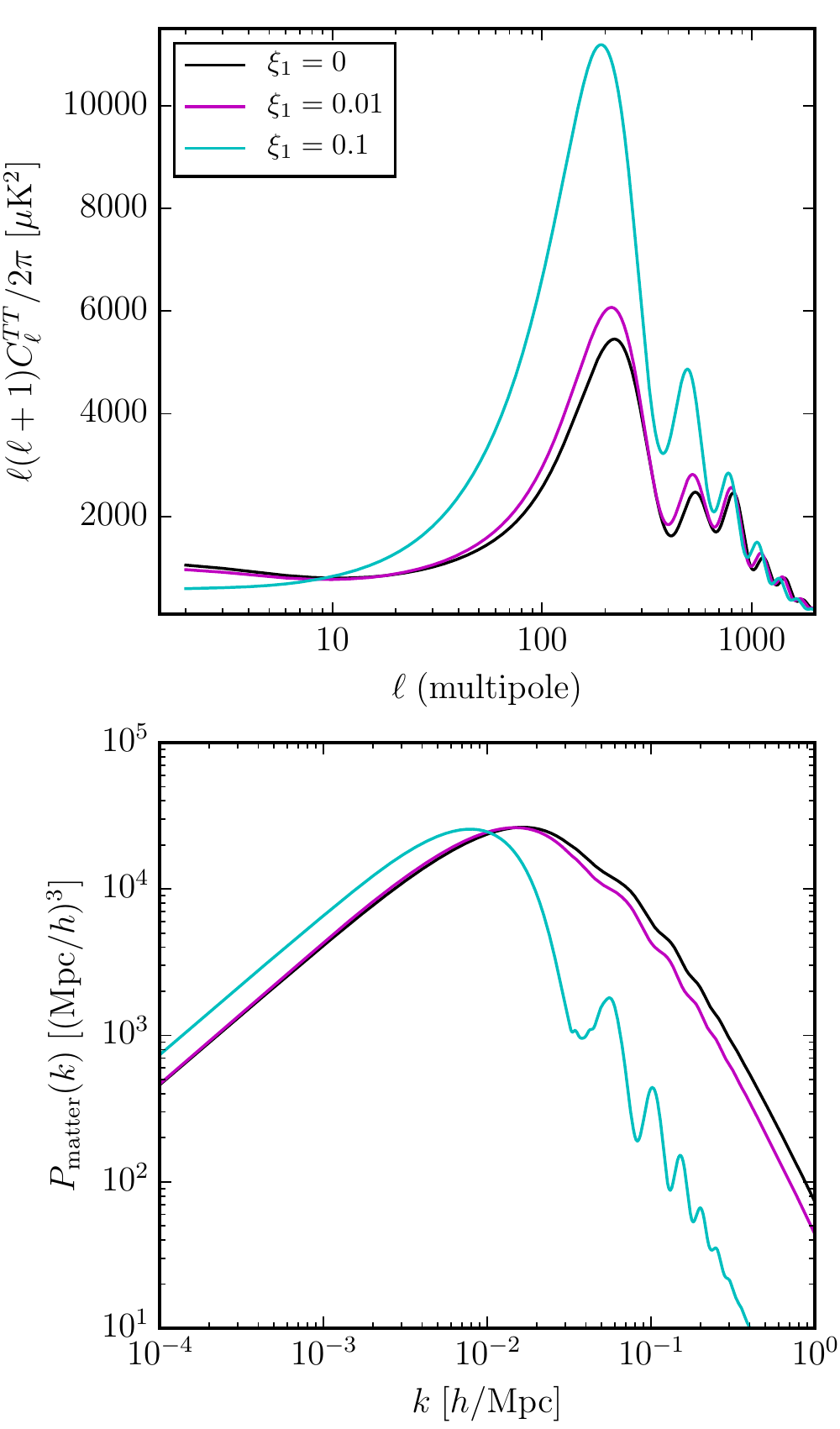}}
 \caption{Plots of the CMB temperature power spectrum (upper panel) and matter power spectrum (lower panel)
 for three different coupling parameters $\xi_1$ for model 3.
 The other parameters assume the fiducial values given in Sec.~\ref{Method}, Table \ref{tbl_m3}.
 In particular, $w_\mathrm{DE}=-1.06$.}
\label{model3}
\end{figure}

\begin{figure}
\includegraphics[scale=0.79]{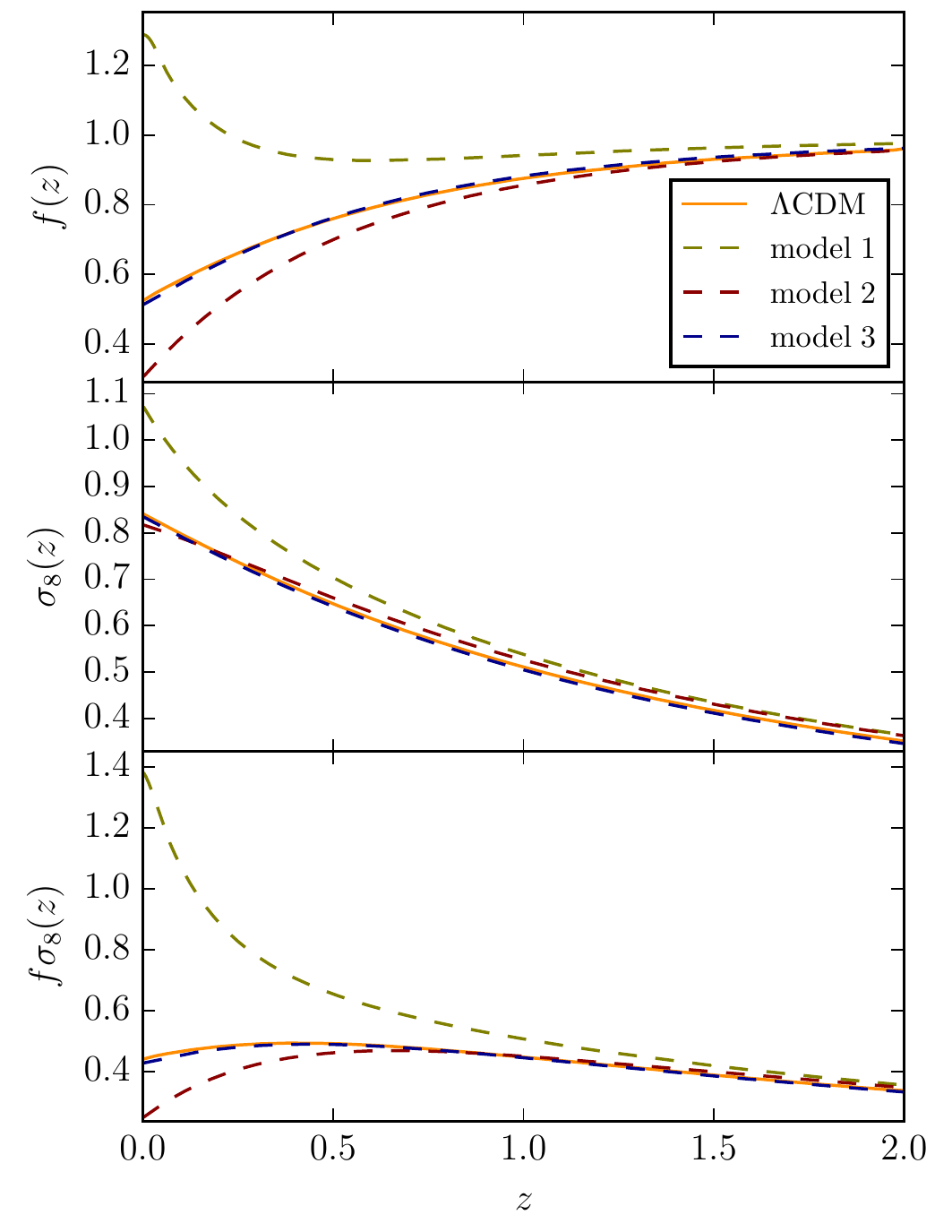}
 \caption{Plots of the growth rate ($f$; top panel), the root mean square matter fluctuations today in linear theory
 at a characteristic length scale of $8~\mathrm{Mpc}/h$ ($\sigma_8$; middle panel), and their product ($f\sigma_8$; bottom panel)
 as a function of redshift ($z$). The orange curve represents $\Lambda$CDM with Planck-like fiducial values for the cosmological parameters
 (see Ref.\ [2]). For models 1, 2, and 3 (green, red, and blue curves, respectively), the fiducial values were taken following
 the best fit values of Ref.\ [17], also shown in Tables III to V of our manuscript.}
\label{fig:growthrate}
\end{figure}

To be able to compare theoretical predictions from the different phenomenological models with experiments,
the cosmological perturbations for these models have been calculated in Ref.~\cite{He:2010im}.
In this reference, the linear perturbations are calculated by perturbing the Friedmann-Lema\^itre-Robertson-Walker spacetime and the
energy-momentum tensor of the coupled DM-DE fluid.
First, the background interaction four-vector is given by $Q_{\left(\lambda \right)}^{\nu}= \left[ Q, 0, 0, 0 \right]^T$,
which represents the exchange of energy density only [c.f.~Eq.~\eqref{de_dm}].
The subscript $\lambda$ stands for either DE in the case of models 1 and 2 or DM in the case of model 3.
Then, the perturbed four-vector representing the perturbation in the interaction between the DM and DE
fluids, $\delta Q_{\left( \lambda \right)} ^{\nu}$, can be decomposed into
\begin{align}
 \delta Q^{0}_{(\lambda)} &= \pm \left(-\frac{\psi}{a}Q + \frac{1}{a}\delta Q\right)\,, \nonumber \\
 \delta Q_{p (\lambda)} &= \left. Q_{p (\lambda)}^I \right|_t + Q^{0}_{(\lambda)}v_t\,.
\end{align}
The $\pm$ sign refers to DM or DE respectively; $\delta Q_{p (\lambda)}$ is the potential of the perturbed energy-momentum transfer
$\delta Q^{i}_{(\lambda)}$; $\left. Q_{p (\lambda)}^I \right|_t$ is the external non-gravitational force density;
and $v_t$ is the average velocity of the energy transfer.
Since we have a stationary energy transfer, we only consider inertial drag effects, so $\left. Q_{p (\lambda)}^I \right|_t$ and $v_t$ vanish,
which implies that $\delta Q^{i}_{(\lambda)}=0$. One can then evaluate the linear order perturbation equations for DM and DE
(we refer to Ref.~\cite{He:2010im} for more details; see also Ref.~\cite{Honorez:2010rr} for another study of the perturbations in this context).

With the perturbations, one can then compute the CMB temperature angular power spectrum ($C_\ell^{TT}$) and
the matter power spectrum ($P_\mathrm{matter}(k)$).
The corresponding spectra are shown for different values of the coupling constant for each of the models described in Table \ref{ide_m}
in Figs.~\ref{model1} (model 1), \ref{model2} (model 2), and \ref{model3} (model 3).
By computing the perturbations, one can also evaluate the growth rate ($f(z)$) and the root mean square of matter fluctuations today
at a characteristic length scale of $8~\mathrm{Mpc}/h$ ($\sigma_8(z)$) in order to illustrate how the interaction affects the growth of structure.
This is shown in Fig.~\ref{fig:growthrate}.
The plots are generated using a modified version of the \texttt{CAMB} software package \cite{Lewis:1999bs},
which incorporates the physics of interacting DE.
Moreover, the cosmological parameters (excluding the values
for the coupling constants $\xi_2$ and $\xi_1$ for the power spectra) are assumed to take
the fiducial values given in Sec.~\ref{Method}, see Table \ref{tbl_m1}
for Fig.~\ref{model1}, Table \ref{tbl_m2} for Fig.~\ref{model2}, and
Table \ref{tbl_m3} for Fig.~\ref{model3}. 

As noted in Ref.~\cite{He:2010im}, changes in the DE EoS mainly influence the low-$\ell$ angular power spectrum
and can shift the overall amplitude of the matter power spectrum slightly.
For this reason, we only show the changes caused by varying the coupling constant in the power spectra.
From the plots, we see that interacting DE can have effects that are degenerate with changing the DE EoS, but these degeneracies
can be broken by including all the information from both the CMB angular power spectrum and the matter power spectrum today.
Indeed, interacting DE generally changes the size of the CMB acoustic peaks and it affect the amplitude
of $P_\mathrm{matter}$ only at large $k$, which can hardly be mimicked by a different EoS.

Generally, we can see that for large couplings ($\xi_{1,2}=0.1$), the changes in the acoustic peaks of the
power spectra compared to $w$CDM ($\xi_{1,2}=0$) are very pronounced, so large couplings can
be easily ruled out by observations. However, in
general, small couplings introduce more subtle changes
that are harder to be distinguished, and
from previous analyzes, small couplings are preferred by
the observations
\cite{Costa:2013sva,Wang:2016lxa,Costa:2016tpb,Murgia:2016ccp,iDEobsrefPettorino,iDEobsrefSalvatelliMena,iDEobsrefYangXu,iDEobsrefmoreYang,iDEobsrefLi,iDEobsrefNunes,Marcondes:2016reb,An:2017kqu,iDEobsrefothers,Abdalla:2014cla},
although with small significance.
Although subtle, these changes behave differently depending on the model chosen,
so it is important to understand how each model affects the power spectra.

For model 1, let us first note that the interacting constant must be negative \cite{Costa:2016tpb}.
This means that there is an energy flow from DM to DE and that the DM energy density is higher in the past compared to $\Lambda$CDM.
As we can see in Fig.~\ref{model1}, having more DM in the past leads to an overall suppression of the angular power spectrum
together with a small change in the low-$\ell$ behaviour. This extra amount of DM influences the evolution of matter perturbations,
leading to more structure formation, as it can be seen in Fig.~\ref{fig:growthrate}. The growth rate is always higher than in
$\Lambda$CDM, especially at low redshifts when DE starts to dominate at which point these effects become more pronounced.
This also affects $\sigma_8 (z)$ at low-redshifts. As a result, there is an enhancement in the amplitude of the matter power spectrum,
mainly on scales smaller than the turn-around point due to the fact that the matter-radiation equality happened earlier in the evolution of
the universe, thus decreasing the amount of damping of the small-scale modes during radiation domination.

For models 2 and 3 (see Figs.~\ref{model2} and \ref{model3}), one has a positive coupling constant,
meaning that there is less DM in the past in comparison with $\Lambda$CDM, as DE gets converted into DM as time evolves.
This leads to an enhancement of the overall amplitude of the peaks in the angular power spectrum caused by shallower DM gravitational
potentials, leading to the elimination of baryon loading. This affects the evolution of the matter perturbations, which can be seen by a
slightly smaller growth rate, for model 2 (see Fig.~\ref{fig:growthrate}). This change is very subtle for model 3, since the density of DM
does not depend so strongly on the energy density of DE, but its deviation from the standard dust behaviour depends mainly on the
interaction \cite{He:2010im}. The effect on the matter power spectrum is a change in the turn-around position,
caused by the matter-radiation equality being shifted to lower redshifts in comparison to $\Lambda$CDM, which in turn leads to damping on small scales.
Model 3 presents a more pronounced effect for a large coupling, as we can see in Fig.~\ref{model3}.
This choice of large coupling is not realistic and excluded by observations, but it serves to illustrate the effects that
the interaction has in the power spectrum and the possible degeneracy between the coupling and $\Omega_{\mathrm{c}}$. 

As we saw above, changes in the power spectra from the interaction between DM and DE can be mimicked by changing $\omega_\mathrm{c}$
or by a different EoS of DE, showing again the degeneracy between the coupling constant, $\omega_\mathrm{c}$, and the EoS of DE.
In order to measure a small coupling between DM and DE and to improve the constraints on the cosmological parameters, we need
future generations of cosmological observations and complementary observations that can break the degeneracy
between the interacting DE parameters and the DM energy density. Different observations
probe DE and other cosmological parameters in different ways. For this reason, in
what follows, we explore the combination of CMB and BAO observations as a mean to break degeneracies.
We will show that some of those degeneracies can indeed be broken by combining CMB and BAO measurements, but to fully break the degeneracies,
growth of structure measurements like weak lensing and galaxy clustering (that would measure, e.g., $f\sigma_8$ as shown in Fig.~\ref{fig:growthrate})
should be included. We keep this for follow-up work. For a first theoretical analysis of the growth of structures in
interacting DE models, see Refs.~\cite{CalderaCabral:2009ja,Marcondes:2016reb}.

\section{Methodology\label{Method}}

\subsection{Information from galaxy surveys: baryon acoustic oscillation}

The baryon acoustic oscillation (BAO) is an important observable currently used to constrain the cosmological parameters
more efficiently in combination with other probes such as the CMB. The information stored in the BAO peaks present in the matter power
spectrum can be used to determine the Hubble parameter $H(z)$ and the angular diameter distance $D_A(z)$ as a function of the redshift,
which subsequently allows us to calculate the DE parameters. Let us first define the observed power spectrum in
redshift space using a particular reference cosmology (in our case, $\Lambda$CDM), which differs from the true cosmology
(for details about this methodology, see Ref.~\cite{Seo:2003pu}), as follows,
\begin{align}
P_\mathrm{obs}(k^\mathrm{(ref)}_{\perp}, k^\mathrm{(ref)}_{\parallel})=&~\left(\frac{D_A^\mathrm{(ref)}(z)}{D_A(z)}\right)^2
 \left(\frac{H(z)}{H^\mathrm{(ref)}(z)}\right) \nonumber \\
 &\times P_{g}(k_\perp, k_\parallel)+P_\mathrm{shot}\,, \label{p_obs}
\end{align}
where $P_\mathrm{shot}$ is the unknown Poisson shot noise.
The Hubble parameter $H(z)$ and angular diameter distance $D_A(z)$ values in the reference cosmology ($\Lambda$CDM) are distinguished
from the values in the true cosmology by the superscript `(ref)'.

The angular diameter distance is given by
\begin{equation}
\label{da2}
D_A(z)= \frac{c}{1+z}\int_{0}^{z} \frac{\mathrm{d}z}{H(z)}\,,
\end{equation}
hence it depends on the evolution of the Hubble parameter.
We can write $H(z)$ as a function of the DE and DM parameters, knowing
that it is related to the DE and DM densities through the Friedmann equation,
\begin{equation}
\label {hubble_parameter}
H(z)^2 = \frac{8\pi G}{3} \left[\rho_\mathrm{DE}(z)+\rho_\mathrm{DM}(z)+\rho_\mathrm{b}(z) \right]\,,
\end{equation}
where the evolution of the different energy densities depend on the model chosen
as seen in Sec.~\ref{models} [cf.~Eqs.~\eqref{eq:rho12} and \eqref{eq:rho3}].

The wavenumbers across and along the line of sight in the true cosmology are denoted by $k_{\perp}$ and $k_{\parallel}$,
and they are related to the ones in the reference cosmology by
$k^\mathrm{(ref)}_{\perp} = k_{\perp} D_A(z)/D_A^\mathrm{(ref)}(z)$ and $k^\mathrm{(ref)}_{\parallel} = k_{\parallel}H^\mathrm{(ref)}(z)/H(z)$.
The galaxy power spectrum, $P_g$, can be written as follows:
\begin{align}
\label{p_g}
P_{g}(k^\mathrm{(ref)}_{\perp}, k^\mathrm{(ref)}_{\parallel})=&~b^2(z)\left(1+\beta\mu^2\right)^2 \left(\frac{G(z)}{G(z=0)}\right)^2 \nonumber \\
 &\times P_{\mathrm{matter},\,z=0}(k) e^ {-k^2 \mu^2 \sigma_r^2}\,.
\end{align}
In the equation above, we defined $\mu\equiv\mathbf{k}\cdot\hat{\mathbf{r}}/k$, where $\hat{\mathbf{r}}$ is the unit vector along the line of sight.
The exponential damping factor is due to redshift uncertainties ($\sigma_z$), where $\sigma_r\equiv c\sigma_z/H(z)$.
Also, $G(z)$, $\beta(z)$, and $b(z)$ are the growth function, the linear redshift space distortion parameter, and the linear galaxy bias,
respectively, which are related through the definition $\beta(z)\equiv f/b(z)$.
The linear matter power spectrum, $P_{\mathrm{matter},\,z=0}(k)$, as well as the growth rate, $f$,
are generated using a modified version of \texttt{CAMB} to account for the physics of interacting DE.
The effect of the interaction in these models was described in the previous section.

The above provides the necessary information to perform a Fisher matrix forecast for future BAO experiments.
The Fisher matrix formalism has become the standard method for predicting the precision with which various cosmological
parameters can be extracted from future data. The advantage of it relies on the fact that it is a fast approach
and generally returns accurate estimates for the parameter errors from the derivatives of the observables with
respect to the model parameters around the best fit value.
We note, though, that it is not always justified to use the Fisher matrix approach as opposed
to a Monte Carlo Markov Chain (MCMC) posterior likelihood estimation method (see, e.g., Ref.~\cite{Wolz:2012sr}).
This is especially true when one does not know whether the cosmological parameters of the given theoretical model
will be Gaussian or not for a given set of cosmological data. This is why older studies have preferred an MCMC
approach \cite{Martinelli:2010rt,DeBernardis:2011iw}, but these papers have shown that the estimated likelihood contours for cosmological parameters
of phenomenological interacting DE could be well-approximated by Gaussian ellipses.
Furthermore, many MCMC analyzes with current data have shown similar Gaussian-like likelihood curves.
Hence, we believe that the Fisher matrix approach is well justified in this case, though we must keep in mind
that the constraints found are probably lower bounds on the marginalized errors (i.e., it is the best-case scenario).

For the matter power spectrum obtained from galaxy
surveys, the Fisher matrix is given by (see Ref.~\cite{Tegmark:1997rp})
\begin{align}
F_{ij}=&~\int_{-1}^1\int_{k_\mathrm{min}}^{k_\mathrm{max}}\frac{\partial\ln P_g(k, \mu)}{\partial p_i}\frac{\partial\ln P_g(k,\mu)}{\partial p_j} \nonumber \\
 &\times V_\mathrm{eff}(k,\mu)\,\frac{2\pi k^2~\mathrm{d}k\,\mathrm{d}\mu} {2(2\pi)^3}\,,
\label{fisher_gal2}
\end{align}
where $p_i$ and $p_j$ are elements of the set of parameters for the given cosmological model.
The effective volume of the survey, $V_{\textrm{eff}}$, can be written, for a constant comoving number density ($\bar{n}$), as
\begin{equation}
\label {veff}
V_{\textrm{eff}}(k,\mu)=\left[\frac{\bar{n}P_g(k,\mu)}{1+\bar{n}P_g(k,\mu)}\right]^2V_\mathrm{survey}\,.
\end{equation}
In this paper, we present the expected cosmological implications of the BAO measurements for a Euclid-like survey
(for specifications of the Euclid survey, see, for example, Ref.~\cite{Amendola:2012ys} and references therein). We assume an area of
$15\,000~\mathrm{deg}^2$, a redshift accuracy of $\sigma_z/(1+z)=0.001$, and a redshift range $0.5\leq z\leq 2.1$.

We then take 15 redshift bins of width $\Delta z=0.1$ centered on
$z_i$. The set of parameters of interest to obtain constraints on the dark sector is
$\mathcal{P}=\{\omega_\mathrm{b}\equiv h^2\Omega_\mathrm{b},\omega_\mathrm{c},h,H(z_i),D_A(z_i),G(z_i),\beta(z_i),P^i_\mathrm{shot}\}$.
For a given redshift slice, the parameters $H(z_i)$, $D_A(z_i)$,
$G(z_i)$, $\beta(z_i)$, and $P^i_\mathrm{shot}$ are estimated
simultaneously with $\omega_\mathrm{b}$, $\omega_\mathrm{c}$,
and $h$ and according to the assumed fiducial values of a set of
cosmological parameters for each considered model. The total number
of parameters is $5N+3$ for a BAO survey divided in $N$ redshift
bins. The derivatives of the observable $P_g$ with respect to the model
parameters in Eq.~\eqref{fisher_gal2} are then evaluated at the fiducial
values, which we take to be the best-fit values of
Ref.~\cite{Costa:2016tpb} for each considered interacting DE model.
Finally, we must derive the errors on
$H(z)$ and $D_A(z)$ to later propagate them into the desired dark
sector parameters for the interacting DE models.

After marginalizing the Fisher matrix defined above over
$G(z_i)$, $\beta(z_i)$, and $P^i_\mathrm{shot}$, a sub-matrix is then calculated as follows,
\begin{equation}
\label{fisher_gal3}
F^{\mathrm{DE}}_{mn}=\sum_{\alpha,\,\beta}\frac{\partial p_\alpha}{\partial q_m}F^\mathrm{(sub)}_{\alpha\beta}\frac{\partial p_\beta}{\partial q_n}\,,
\end{equation}
where $p_\alpha,p_\beta\in\mathcal{P}\setminus\{G(z_i),\beta(z_i),P^i_\mathrm{shot}\}$ and $q_m,q_n\in\mathcal{Q}$,
the latter being the final set of parameters defined as
$\mathcal{Q}=\{ \omega_\mathrm{b}, \omega_\mathrm{c}, h, w_\mathrm{DE}, \xi_2 \}$ for models 1 and 2
and $\mathcal{Q}=\{ \omega_\mathrm{b}, \omega_\mathrm{c}, h, w_\mathrm{DE}, \xi_1 \}$ for model 3.

The constraints on the dark sector parameters are then determined by how well the survey is able to estimate the values of $H(z)$ and $D_A(z)$.

\subsection{Information from CMB}

In the context of cosmological parameters forecast, we use the CMB information as a second probe to test the ability of
future surveys to constrain a possible interaction in the dark sector and possibly to distinguish between the different
interacting models described previously and the $\Lambda$CDM model. We use the modified \texttt{CAMB} software package \cite{Lewis:1999bs}
to generate the numerical power spectra ($C_\ell^{TT},C_\ell^{EE},C_\ell^{TE}$) for our cosmological models with $\ell\leq 3000$.
We do not consider primordial $B$-modes (i.e., we assume a vanishing primordial tensor power spectrum)
or CMB lensing in the analysis. The latter is justified by the fact
that the \texttt{HALOFIT} \cite{Smith:2002dz} non-linear implementation present in \texttt{CAMB} has only been tested against N-body simulations
for $\Lambda$CDM cosmologies and the non-linear structure evolution starts to affect the lensing signal already at
$\ell>400$ (see Ref.~\cite{lensingref} for studies of CMB lensing and of the non-linear regime in coupled DE cosmologies).
We then construct the Fisher matrix for the CMB temperature anisotropy and polarization as follows (see Ref.~\cite{Zaldarriaga:1996xe}),
\begin{equation}
\label{fisher_unl}
F_{ij}=\sum_\ell\sum_{X,\,Y}\frac{\partial C^{X}_\ell}{\partial p_i}(\mathrm{Cov}_{\ell}^{-1})_{XY}\frac{\partial C^{Y}_\ell}{\partial p_j}\,,
\end{equation}
where $C^{X}_\ell$ represents the power in the $\ell$-th multipole, and where
$X$ stands for $TT$ (temperature), $EE$ ($E$-mode polarization), and $TE$ (temperature and $E$-mode polarization cross-correlation).
The covariance matrix is given by
\begin{equation}
 [\mathrm{Cov}_\ell]=\frac{2}{(2\ell+1)f_\mathrm{sky}}
 \left[\begin{array}{rrrr}
\Xi_\ell^{TTTT} & \Xi^{TTEE}_\ell & \Xi^{TTTE}_\ell   \\
\Xi^{TTEE}_\ell & \Xi^{EEEE}_\ell & \Xi^{EETE}_\ell   \\
\Xi^{TTTE}_\ell & \Xi^{EETE}_\ell & \Xi^{TETE}_\ell
\end{array}\right]\,,
\label{cov_array}
\end{equation}
and the elements of the matrix are given in Appendix \ref{appendixA}.

\begin{table}[ht]
\centering
\caption{Advanced ACT \cite{Calabrese:2014gwa} specifications with $f_\mathrm{sky}=0.5$.
The frequency of the detector, the beam resolution ($\theta_\mathrm{beam}$),
and the map noise ($\sigma_T$) are given in the three columns.\label{ATC}}
\begin{ruledtabular}
\begin{tabular}{ccc}
 Frequency [GHz] & $\theta_\mathrm{beam}$ & $\sigma_T [\mu\mathrm{K}$-$\mathrm{arcmin}]$ \\
 \hline
 90   & $2.2^\prime$ & 7.8  \\
 150  & $1.3^\prime$ & 6.9  \\
 230  & $0.9^\prime$ & 25
\end{tabular}
\end{ruledtabular}
\end{table}

In the near future, CMB surveys will continue to improve, especially ground-based instruments
designed to measure polarization. The Advanced Atacama Cosmology telescope (AdvACT) \cite{Calabrese:2014gwa} is expected
to obtain precise measurements of the CMB small-scale polarization, enabling us to tackle a wide range
of cosmological physics. In particular, it will tighten the constraints on the cosmological parameters of alternative models to $\Lambda$CDM.
The instrumental setup of AdvACT is outlined in Table \ref{ATC}. This is the information that we incorporate in our Fisher matrix analysis
to obtain the CMB forecast (see Appendix \ref{appendixA} for details about how noise is handled).

\section{Results and discussion\label{Results}}

Following the methodology described in the previous section, we compute the Fisher matrices
for the three interacting DE models presented in Sec.~\ref{models}
considering a Euclid-like future BAO survey and an AdvACT-like future CMB experiment.
We also consider the combination of BAO and CMB future measurements.
Assuming that the probes are uncorrelated, one can add the Fisher matrices as follows \cite{Coe:2009xf}
\begin{equation}
\label {fisher_unl_len_euclid}
F^{\textrm{total}}_{ij} = F^{\textrm{BAO}}_{ij} + F^{\textrm{CMB}}_{ij}.
\end{equation}

\begin{table}[ht]
\centering
\caption{Marginalized errors (68\% C.\,L.) for the DE and DM parameters for model 1.
The forecasted errors are given assuming data from Advanced ACT (CMB) and Euclid (BAO) alone,
and the last column gives the combined forecast.
Recall that we define $\omega_\mathrm{b}\equiv h^2\Omega_\mathrm{b}$, $\omega_\mathrm{c}\equiv h^2\Omega_\mathrm{c}$,
and $h=H_0/(100~\mathrm{km/s/Mpc})$.\label{tbl_m1}}
\begin{ruledtabular}
\begin{tabular}{ccccc}
Parameter & Fiducial & AdvACT & Euclid & AdvACT + Euclid \\
          & value    & (CMB)  & (BAO)  &                 \\
\hline
$\omega_\mathrm{b}$  & 0.02224 & 3.86e-05 & 0.00028 & 3.69e-05 \\
$\omega_\mathrm{c}$  & 0.08725 & 0.017    & 0.0017  & 0.00053  \\
$h$                  & 0.6845  & 0.0079   & 0.0055  & 0.0014   \\
$w_\mathrm{DE}$      & -0.9434 & 0.028    & 0.026   & 0.0044   \\
$\xi_2$              & -0.0929 & 0.045    & 0.0037  & 0.0019
\end{tabular}
\end{ruledtabular}
\end{table}

\begin{figure*}
{\includegraphics[scale=0.706]{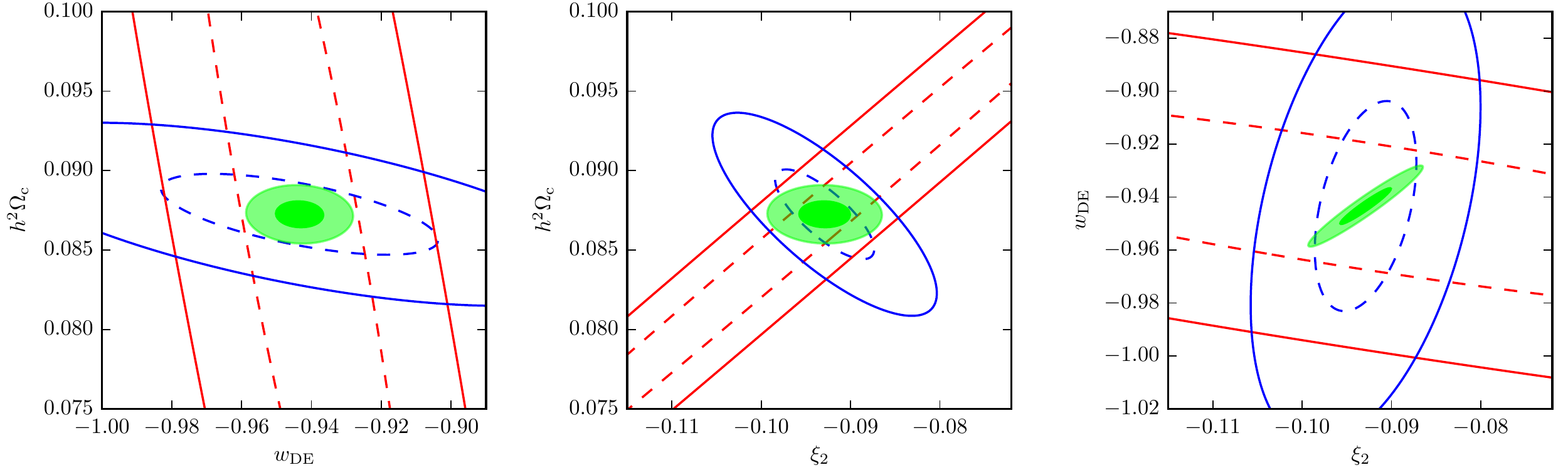}}
 \caption{Fisher forecast contours for model 1 with CMB and BAO information using AdvACT (red curves) and Euclid (blue curves) experimental setups,
 respectively. The dashed curves represent 68\% C.\,L. and the solid curves represent 99.9\% C.\,L.
 The combined contours are shown by the green filled ellipses.
 Similarly, the darker ellipses represent 68\% C.\,L. and the fainter ones represent 99.9\% C.\,L.
 See Table \ref{tbl_m1} for numerical values.}
\label{fisher_modelo1}
\end{figure*}

\begin{figure*}
\includegraphics[scale=0.75]{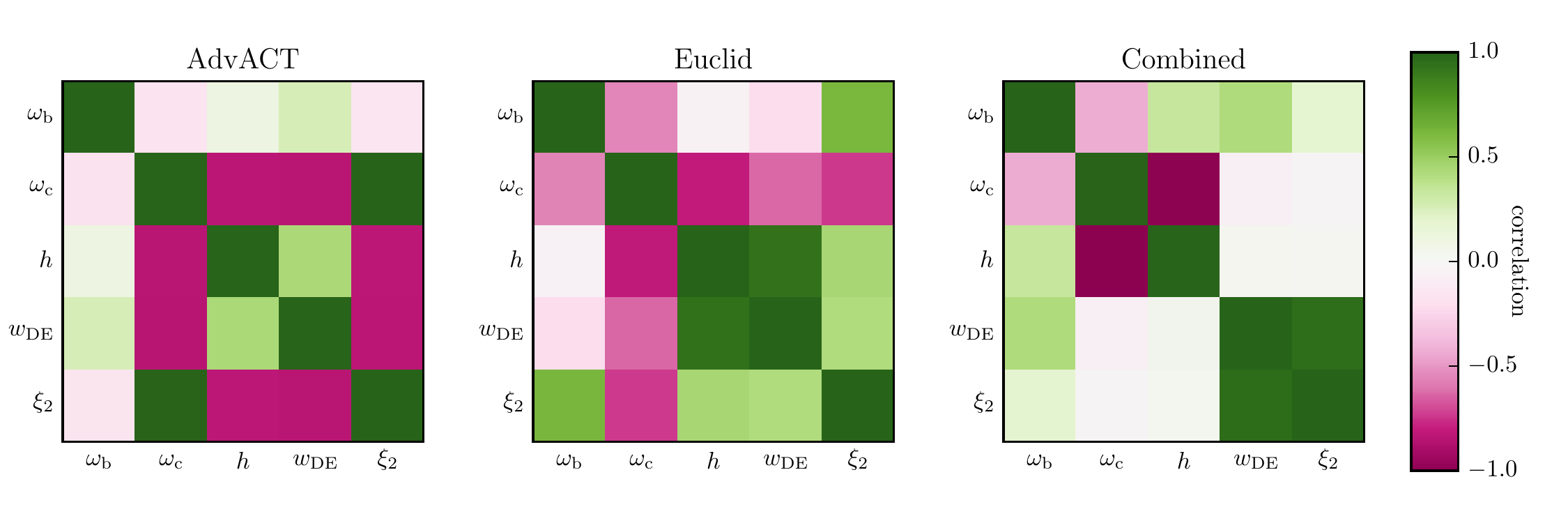}
 \caption{Correlation matrix computed according to Eq.~\eqref{eq:correlationmatrix} for AdvACT (left), Euclid (center), and their combination (right)
 for model 1. The color in each cell indicates the correlation between two model parameters, ranging from 0 (completely independent)
 to $\pm 1$ (completely \mbox{(anti-)correlated}).}
\label{fisher_modelm1}
\end{figure*}

We first show the result for model 1 in Table \ref{tbl_m1},
where the different columns represent the cosmological parameter, its fiducial value,
and the 68\% confidence level (C.\,L.) constraints that would result from AdvACT, Euclid, and the combination of AdvACT and Euclid, respectively.
We notice that the marginalized error for the DE EoS improves drastically for the combined analysis, being $\sigma(w_\mathrm{DE})= 0.026$ for Euclid,
$\sigma(w_\mathrm{DE})= 0.028$ for AdvACT, and $\sigma(w_\mathrm{DE})= 0.0044$ for their combination: an improvement by a factor of $\sim 6$
when compared with each individual probe. The constraint on the DM density improves by a factor of $\sim 3$ for the combined
analysis ($\sigma(\omega_\mathrm{c})= 0.00053$), compared with Euclid alone ($\sigma(\omega_\mathrm{c})= 0.0017$). A similar improvement
occurs for the coupling constant, where we find $\sigma(\xi_2)= 0.0037$ for Euclid alone and $\sigma(\xi_2)= 0.0019$ for $\mathrm{AdvACT + Euclid}$.
Such a stringent constraint would exclude the null interaction corresponding to $w$CDM with high confidence
given that the contours from global fits would be centered on values close to the fiducial values used in this analysis.

Present constraints on $\omega_\mathrm{c}$, $w_\mathrm{DE}$, and $\xi_2$ for a combination of probes
(Planck+BAO+SNIa+H0; see Ref.~\cite{Costa:2016tpb}) are found to be
$\omega_\mathrm{c}=0.0792^{+0.0348}_{-0.0166}$, $w_\mathrm{DE}=-0.9191^{+0.0222}_{-0.0839}$, and $\xi_2=-0.1107^{+0.085}_{-0.0506}$.
The fact that our forecast suggests that future surveys will greatly improve these constraints can be seen from
the confidence regions of cosmological parameters related to the dark sector.
In Fig.~\ref{fisher_modelo1}, we plot the marginalized confidence ellipses at 1$\,\sigma$ and 3$\,\sigma$
for the combinations of $\omega_\mathrm{c}$, $\xi_2$, and $w_\mathrm{DE}$
for AdvACT (red), Euclid (blue), and $\mathrm{AdvACT + Euclid}$ (green) for model 1.
The constraints on the cosmological parameters are affected by the degeneracies present among them.
In order to assess these degeneracies and to see how introducing new observations like BAO from Euclid can affect them,
we introduce the correlation matrix $\rho_{ij}$, which measures the correlation between two parameters $p_i$ and $p_j$.
It is given by
\begin{equation}
\label{eq:correlationmatrix}
 \rho_{ij}=\frac{\mathrm{Cov}_{ij}}{\sqrt{\mathrm{Cov}_{ii}\mathrm{Cov}_{jj}}}\,,
\end{equation}
where $\mathbf{Cov}\equiv\mathbf{F}^{-1}$, and $\mathbf{F}$ is either the CMB, BAO, or total Fisher matrix.
The correlation coefficient for $i \neq j$ ranges from 0 (the two parameters are completely independent)
to $\pm 1$ (the parameters are completely \mbox{(anti-)correlated}).
The correlation matrix is depicted in Fig.~\ref{fisher_modelm1} for model 1 where white, dark magenta, and dark green are equivalent to $\rho_{ij}=0$,
$-1$, and $1$, respectively.

\begin{table}[ht]
\centering
\caption{Marginalized errors (68\% C.\,L.) for the DE and DM parameters for model 2.
See the caption of Table \ref{tbl_m1} for more details.\label{tbl_m2}}
\begin{ruledtabular}
\begin{tabular}{ccccccc}
Parameter & Fiducial & AdvACT & Euclid & AdvACT + Euclid \\
          & value    & (CMB)  & (BAO)  &                 \\
\hline
$\omega_\mathrm{b}$       & 0.02229 & 3.85e-05 & 0.00022 & 3.76e-05 \\
$\omega_\mathrm{c}$       & 0.1314  & 0.015    & 0.0030  & 0.0010   \\
$h$                       & 0.6876  & 0.075    & 0.0068  & 0.0019   \\
$w_\mathrm{DE}$           & -1.087  & 0.19     & 0.033   & 0.0053   \\
$\xi_2$                   & 0.03798 & 0.055    & 0.0055  & 0.0031
\end{tabular}
\end{ruledtabular}
\end{table}

\begin{figure*}
{\includegraphics[scale=0.706]{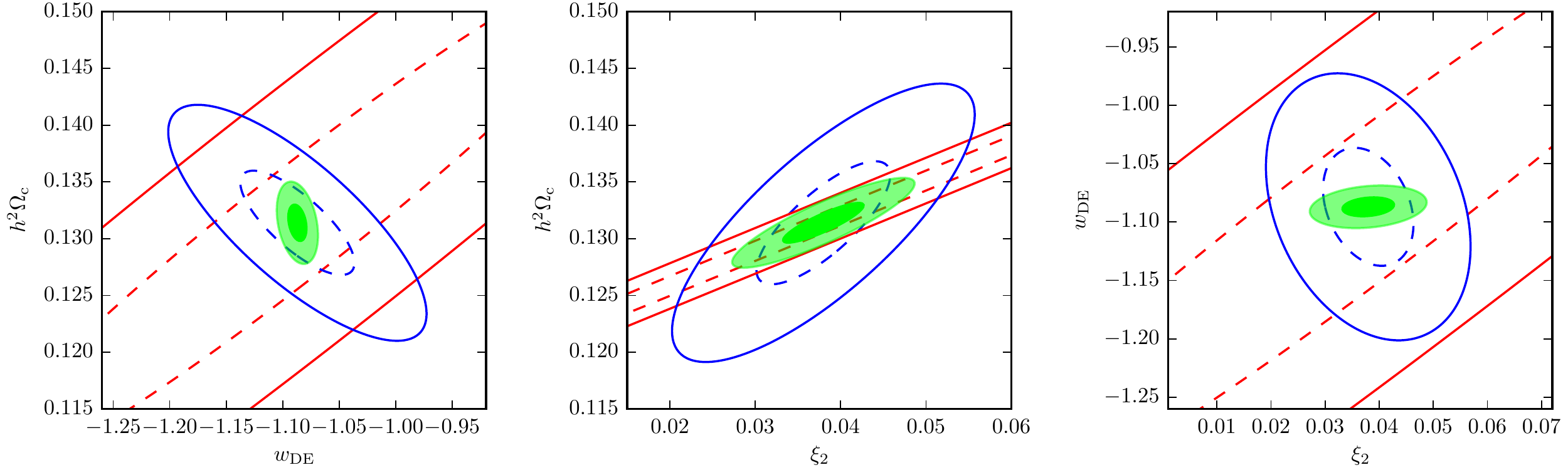}}
 \caption{Fisher forecast contours for model 2.
 The convention used to denote the various cases is described in Fig.~\ref{fisher_modelo1}.
 See also Table \ref{tbl_m2} for numerical values.}
\label{fisher_modelo2}
\end{figure*}

\begin{figure*}
{\includegraphics[scale=0.75]{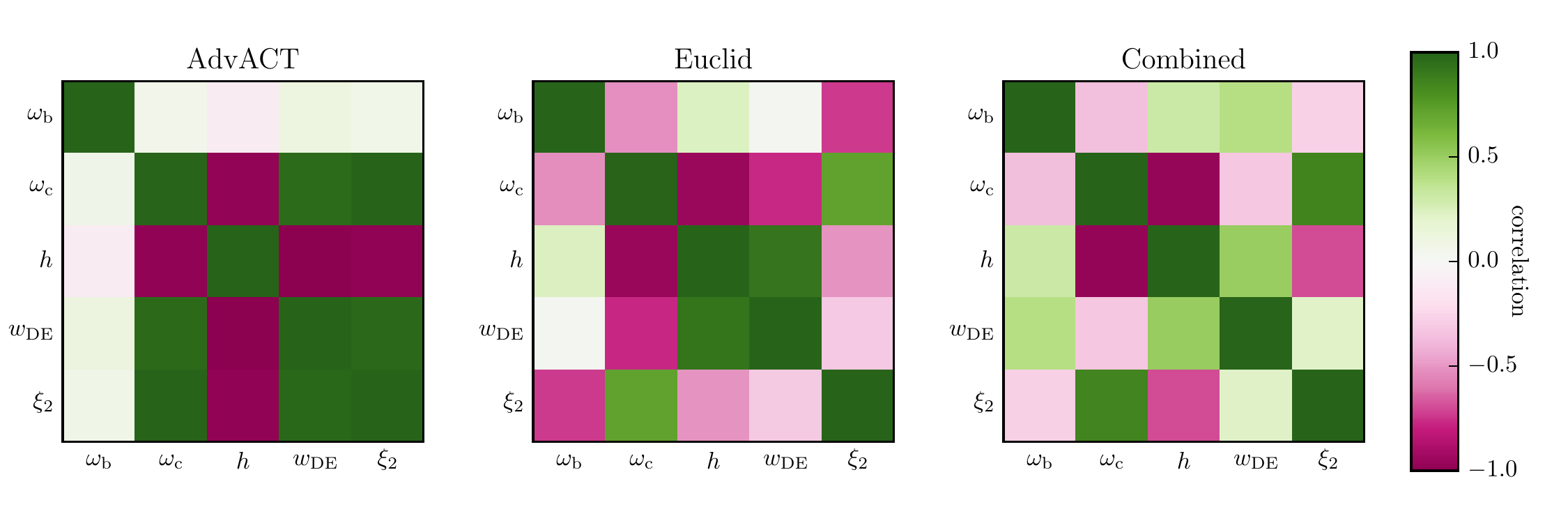}}
 \caption{Correlation matrix computed according to Eq.~\eqref{eq:correlationmatrix} for AdvACT (left), Euclid (center), and their combination (right)
 for model 2. The color in each cell indicates the correlation between two model parameters, ranging from 0 (completely independent)
 to $\pm 1$ (completely \mbox{(anti-)correlated}).}
\label{fisher_modelm2}
\end{figure*}

For model 1, when only CMB information is provided, the dark sector parameters and $\omega_\mathrm{c}$ are very \mbox{(anti-)correlated}
as pointed in Sec.~\ref{models}, which can be seen by visual inspection of Fig.~\ref{fisher_modelo1}.
We can see that the correlation between $w_\mathrm{DE}$ and $\xi_2$ is $\gtrsim 0.8$ (in absolute value)
and it is very large ($\approx 1$) between $\omega_\mathrm{c}$ and $\xi_2$ as well (see Fig.~\ref{fisher_modelm1}).
These degeneracies are considerably weakened when BAO information is added.
For instance, the correlations between $\omega_\mathrm{c}$ and $\xi_2$
and between $\omega_\mathrm{c}$ and $w_\mathrm{DE}$ are reduced to $\approx-0.026$ and $\approx-0.059$, respectively,
for the combined forecast ($\mathrm{AdvACT + Euclid}$).
The correlation between $w_\mathrm{DE}$ and $\xi_2$ changes sign in comparison with CMB alone, and the level of
degeneracy between these parameters is only mildly alleviated. This happens since the degeneracy is present between
the three parameters $w_\mathrm{DE}$, $\xi_2$ and $\omega_\mathrm{c}$, and BAO only helps
constraining one of them, leaving some degeneracy among the other two (or combinations of such parameters).
In summary, these results show that BAO has the power to break some degeneracies, as we can see by the tighter constraints and
the milder correlations encountered, although there remains some degeneracies among parameters.

\begin{figure*}
{\includegraphics[scale=0.706]{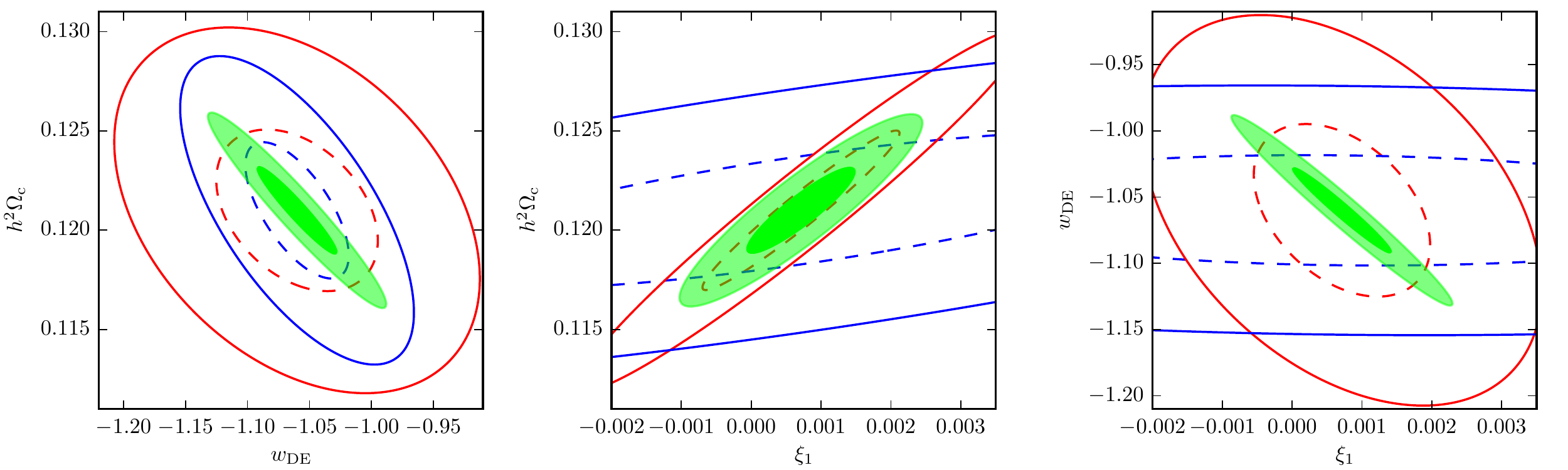}}
 \caption{Fisher forecast contours for model 3.
 The convention used to denote the various cases is described in Fig.~\ref{fisher_modelo1}.
 See also Table \ref{tbl_m3} for numerical values.}
\label{fisher_modelo3}
\end{figure*}

\begin{figure*}
{\includegraphics[scale=0.75]{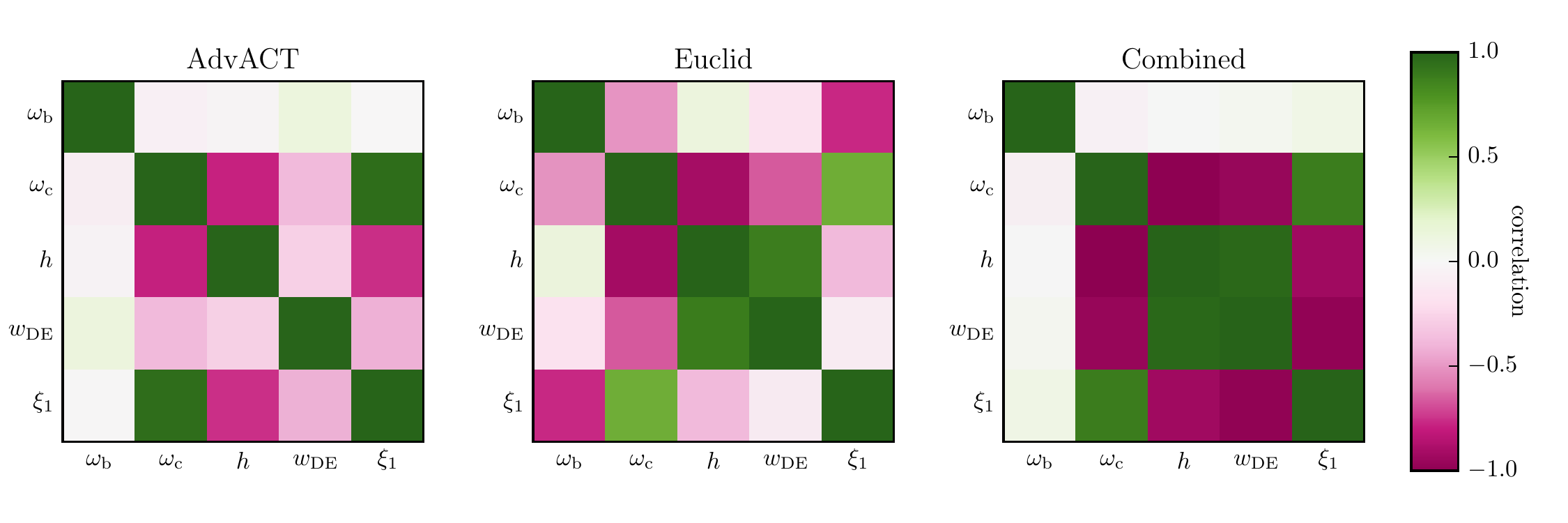}}
 \caption{Correlation matrix computed according to Eq.~\eqref{eq:correlationmatrix} for AdvACT (left), Euclid (center), and their combination (right)
 for model 3. The color in each cell indicates the correlation between two model parameters, ranging from 0 (completely independent)
 to $\pm 1$ (completely \mbox{(anti-)correlated}).}
\label{fisher_modelm3}
\end{figure*}

Similar results are found for model 2 as it can be seen in Table \ref{tbl_m2} and Figs.~\ref{fisher_modelo2} and \ref{fisher_modelm2}.
The combined forecast leads to stringent constraints on $\omega_\mathrm{c}$, $w_\mathrm{DE}$, and $\xi_2$, the latter being $\sigma(\xi_2)=0.00310$.
It was claimed by Ref.~\cite{Murgia:2016ccp} that an energy flow from DE to DM ($\xi_2 > 0$), resulting in a non-zero coupling
between the two dark components where DE decays into DM, is in better agreement with present cosmological data. A vanishing
interaction is also excluded by Ref.~\cite{Costa:2016tpb} with $\xi_2=0.02047^{+0.00565}_{-0.00667}$ (the
errors are given at 68\% C.\,L.).
The future combination of AdvACT and Euclid-like surveys would be able to improve this constraint by a factor of $\sim 2$,
hence one could potentially distinguish the interacting DE model from $w$CDM (and equivalently from $\Lambda$CDM) by more than 3$\,\sigma$
(and possibly even more than 5$\,\sigma$).
Of course, a proper statistical analysis would have to be done in order to really assess which model is preferred by the data.
From the correlation matrix (Fig.~\ref{fisher_modelm2}), we can see, like in the previous case for model 1,
that all the correlations become milder for the combined analysis,
and this happens for all the degenerate parameters $w_\mathrm{DE}$, $\xi_2$, and $\omega_\mathrm{c}$ almost equally.
This might be the case since, as phantom DE can mimic changes in $\omega_\mathrm{c}$, BAO can only break
the degeneracy between a combination of these parameters (not all of them) as was the case for model 1.

\begin{table}
\centering
\caption{Marginalized errors (68\% C.\,L.) for the DE and DM parameters for model 3.
See the caption of Table \ref{tbl_m1} for more details.\label{tbl_m3}}
\begin{ruledtabular}
\begin{tabular}{ccccc}
Parameter & Fiducial & AdvACT & Euclid & AdvACT + Euclid \\
          & value    & (CMB)  & (BAO)  &                 \\
\hline
$\omega_\mathrm{b}$       & 0.02232   & 3.83e-05 & 0.00021 & 3.59e-05 \\
$\omega_\mathrm{c}$       & 0.121     & 0.0027   & 0.0022  & 0.0014   \\
$h$                       & 0.6793    & 0.018    & 0.0055  & 0.0041   \\
$w_\mathrm{DE}$           & -1.06     & 0.043    & 0.027   & 0.021    \\
$\xi_1$                   & 0.0007127 & 0.00083  & 0.00400 & 0.00046
\end{tabular}
\end{ruledtabular}
\end{table}

Confidence ellipses are shown in Fig.~\ref{fisher_modelo3} for model 3, where we see in the middle and right-hand plots
(as well as in Table \ref{tbl_m3}) that CMB plays an important role in constraining $\xi_1$, revealing
that the interaction between DE and DM is already well constrained by CMB data before the inclusion of information about $H(z)$ evolution.
We can also see that for CMB alone in Fig.~\ref{fisher_modelm3} (see the left-hand plot), the correlations are milder than for models 1 and 2.
This agrees with the fact that for model 3, $w_\mathrm{DE}$ and $\xi_1$ are not degenerate at present times.
In this case though, it appears that BAO does not help a lot to break remaining degeneracies.
Indeed, all three combinations of parameters indicate a large correlation ($|\rho|\gtrsim 0.89$) when combining CMB and BAO probes.

The significance of the constraint on $\xi_1$ is already low at the current observational stage:
Ref.~\cite{Costa:2016tpb} found $\xi_1=0.0007127^{+0.000256}_{-0.000633}$ (68\% C.\,L.) when considering a combination of probes
(Planck, SNIa, BAO, and $H_0$ data). Our forecast indicates that $\mathrm{AdvACT + Euclid}$ would yield $\sigma(\xi_1)=0.00046$
(see Table \ref{tbl_m3}), which would not improve the current best constraint much.
However, one must be careful in doing this comparison because the current best constraint includes information from many other probes
such as local measurements of the Hubble constant today and supernova data, which were not included in the forecast done here.
For the model 3, it thus appears that a combination of other probes would still be needed in order to tighten the present limits.

\section{Summary and conclusions\label{conclusions}}

We focused our study on phenomenological interacting DE models and investigated the impact of two probes on the parameter constraints of such models,
specifically the primary CMB temperature and polarization spectrum and the BAO information from a redshift range of $0.5 \leq z\leq 2.1$.
The advantages of combining different observational probes in constraining cosmological parameters are well known,
and its implication to interacting DE models has been widely addressed. Our motivation was to test the ability of future
experiments to constraint such alternative scenarios and distinguish them from models in which there is no interaction in the dark sector.

For models 1 and 2 where the interaction is proportional to the DE energy density,
stringent constraints were found in the dark sector parameters for the combined probes, especially for the coupling constant.
Specifically, with the choice of fiducial values $\xi_2=-0.0929$ (model 1) and $\xi_2=0.03798$ (model 2),
we predicted 1-$\sigma$ marginalized errors of at best $\sigma(\xi_2)=0.0019$ (model 1) and $\sigma(\xi_2)=0.00310$ (model 2).
Thus, the combination of future CMB and BAO experiments, such as presented here, would probably be able to exclude the null interaction
(corresponding to the $w$CDM model) with a confidence level much greater than 3$\,\sigma$,
although it is important to stress again that a proper statistical analysis will have to be done in order to really
assess which model is preferred by the data.
We also showed that the interacting DE models 1 and 2 are affected by degeneracies, which limits the constraining power
of CMB information alone, but that they can be broken by the addition of Euclid-like BAO measurements,
thus tightening the constraints on the dark sector cosmological parameters and enabling a
deeper discussion of these interacting DE scenarios.

We found that the constraint on the coupling constant for model 3 (where interaction is proportional to the DM energy density)
is not improved as much by the combination of future CMB and BAO
experiments compared with its constraint derived by present datasets. It thus appears that extra information is still necessary for probing this model,
and one could consider introducing the CMB lensing power spectra (possibly including higher order corrections \cite{Marozzi:2016qxl})
and/or the convergence power spectrum from weak cosmic shear.
We leave this investigation for future work.

We end by mentioning that future investigations of interacting DE could also benefit from yet more cosmological probes
such as the redshift dependence of the Alcock-Paczy\'{n}ski effect \cite{Li:2016wbl},
cosmic chronometers (see, e.g., Refs.~\cite{iDEobsrefNunes,cosmicchronometersrefs} and references therein),
21~cm cosmology (see, e.g., Refs.~\cite{21cmexp,21cmtheo} and references therein), gravitational waves \cite{Zhao:2010sz,Caprini:2016qxs}, and more.
These examples of probes may be able to remove even more degeneracies and improve the constraints to another level
and shall be considered in future work.

\vspace{0.5cm}

\begin{acknowledgments}
We thank A.~A.~Costa for sharing his modified version of
\texttt{CAMB} with us. LS and WZ are supported by the National
Natural Science Foundation of China (NSFC) through grants
Nos.~11603020, 11633001, 11173021, 11322324, 11421303, and
11653002. They are also supported by the project of Knowledge
Innovation Program of Chinese Academy of Science, the Fundamental
Research Funds for the Central Universities, and the Strategic
Priority Research Program of the Chinese Academy of Sciences Grant
No.~XDB23010200. EF thanks CNPq (Science without Borders) for
financial support. JQ acknowledges financial support from the
Walter C.~Sumner Memorial Fellowship and from the Vanier Canada
Graduate Scholarship administered by the Natural Sciences and
Engineering Research Council of Canada (NSERC). JQ also wishes to
thank Yi-Fu Cai at USTC for hospitality when this work was
initiated.
\end{acknowledgments}

\vspace{0.5cm}

\appendix

\section{The elements of the CMB covariance matrix\label{appendixA}}

In this appendix, we write down explicitly the elements of the CMB covariant matrix given in Eq.~\eqref{cov_array}:
\begin{align}
\label {cov_l10}
\Xi^{TTTT}_\ell &= \left(\mathcal{C}^{TT}_\ell\right)^2\,, \\
\label {cov_l20}
\Xi^{EEEE}_\ell &= \left(\mathcal{C}^{EE}_\ell\right)^2\,, \\
\label {cov_l30}
\Xi^{TETE}_\ell &= \left(\mathcal{C}^{TE}_\ell\right)^2+\mathcal{C}^{TT}_\ell\mathcal{C}^{EE}_\ell\,, \\
\label {cov_l40}
\Xi^{TTEE}_\ell &= \left(C^{TE}_\ell\right)^2\,, \\
\label {cov_l50}
\Xi^{TTTE}_\ell &= C^{TE}_\ell\mathcal{C}^{TT}_\ell\,, \\
\label {cov_l60}
\Xi^{EETE}_\ell &= C^{TE}_\ell\mathcal{C}^{EE}_\ell\,.
\end{align}
In these equations, we defined $\mathcal{C}^{X}_\ell \equiv C^{X}_\ell + N^{X}_\ell$,
where $N^{TT}_\ell$ and $N^{PP}_\ell$ are the Gaussian random detector noises for temperature and polarization, respectively.
One can express the random noises as follow,
\begin{align}
\label{noise1}
N^{TT}_\ell &= \sum_{\nu}\frac{1}{w_T(\nu)B_\ell(\nu)^{2}}\,, \\
\label{noise2}
N^{PP}_\ell &= \sum_{\nu}\frac{1}{w_P(\nu)B_\ell(\nu)^{2}}\,,
\end{align}
where the sums are over the frequencies ($\nu$) of the detector (see Table \ref{ATC} for AdvACT).
The window function is given by
\begin{equation}
 B_\ell(\nu)^2 = \exp\left[\frac{-\ell(\ell+1)\theta_\mathrm{beam}(\nu)^2}{8\ln 2}\right]\,,
\end{equation}
and the inverse square of the detector noise level for temperature and polarization
determines the weight given to each considered experiment channel (i.e., for each frequency $\nu$) \cite{Eisenstein:1998hr}:
\begin{align}
 w_T(\nu)&=\frac{1}{[\theta_\mathrm{beam}(\nu)\sigma_T(\nu)]^{2}}\,,\\
 w_P(\nu)&=\frac{1}{[\theta_\mathrm{beam}(\nu)\sigma_P(\nu)]^{2}}\,,
\end{align}
where polarization sensitivities are rescaled by a factor of $\sqrt{2}$ with respect to the temperature sensitivities,
i.e., $\sigma_P(\nu)=\sqrt{2}\sigma_T(\nu)$.

\end{document}